# Radiation forces and torque on a rigid elliptical cylinder in acoustical plane progressive and (quasi)standing waves with arbitrary incidence

F.G. Mitri*

A B S T R A C T

Analytical expressions for the axial and transverse acoustic radiation forces as well as the radiation torque per length are derived for a rigid elliptical cylinder placed arbitrarily in the field of in plane progressive, quasi-standing or standing waves. The rigid elliptical cylinder case is important to be considered as a first-order approximation of the behavior of a fluid particle suspended in air, because of the significant acoustic impedance mismatch at the particle's boundary. Based on the partial-wave series expansion method in cylindrical coordinates, non-dimensional acoustic radiation force and torque functions are derived and defined in terms of the scattering coefficients of the elliptic cylinder. A coupled system of linear equations is obtained after applying the Neumann boundary condition for an immovable surface in a non-viscous fluid, and solved numerically by matrix inversion after performing a single numerical integration procedure. Computational results for the non-dimensional force components and torque, showing the transition from the progressive to the (equi-amplitude) standing wave behavior, are performed with particular emphasis on the aspect ratio *a/b*, where *a* and *b* are the semi-axes of the ellipse, the dimensionless size parameter, as well as the angle of incidence ranging from end-on to broadside incidence. The results show that the elliptical geometry has a direct influence on the radiation force and torque, so that the standard theory for circular cylinders (at normal incidence) leads to significant miscalculations when the cylinder cross-section becomes non-circular. Moreover, the elliptical cylinder experiences, in addition to the acoustic radiation force, a radiation torque that vanishes for the circular cylinder case. The application of the formalism presented here may be extended to other 2D surfaces of arbitrary shape, such as Chebyshev cylindrical particles with a small deformation, stadiums (with oval shape), or other non-circular geometries.

## I. INTRODUCTION

Acoustical waves can levitate [1], transport and rotate objects of different shapes and sizes, throughout various media [2] in a non-contact (and thus contamination-free) mode. This process is of paramount importance in tissue engineering [3], crystallography [4], biomaterials [5], fluid dynamics [6] and materials science [7] in the laboratory and microgravity environments [8,9], to name a few applications. The physical mechanism encountered during particle levitation, transport and rotation with continuous waves involve the acoustic scattering, the steady radiation force and torque phenomena, which are closely ertwined. The evaluation of the radiation force and torque requires the integration of time-averaged physical observables, which depend on quadratic terms of the linear (first-order) incident and scattered pressure (or velocity potential) fields [10].

*Email address: F.G.Mitri@ieee.org

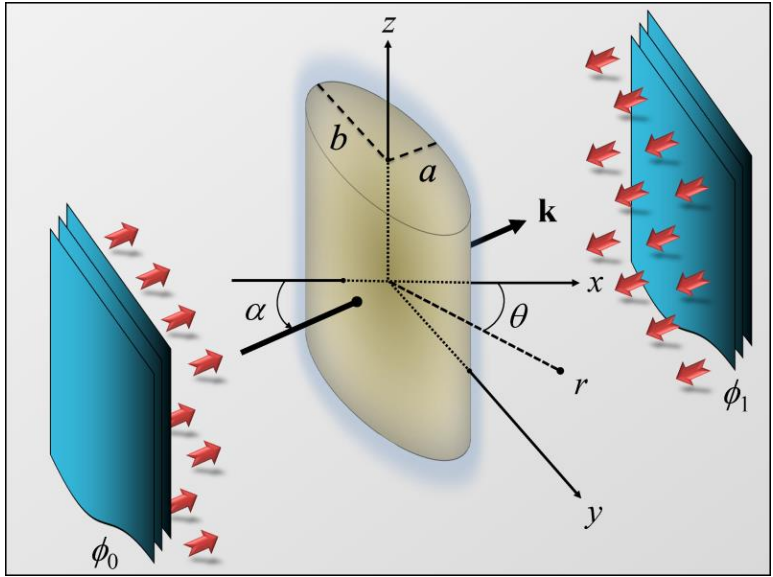

FIG. 1. The schematic describing the interaction of plane (quasi)standing waves with a cylinder of elliptical cross-section with arbitrary incidence. If $\phi_1 = 0$, the incident field corresponds to plane progressive waves. The semi-axes of the ellipse are denoted by $a$ and $b$, respectively. The cylindrical coordinate system $(r, \theta, z)$ is referenced to the center of the elliptical cylinder, where the $z$-axis is perpendicular to the plane of the figure.

In practice, there is no particular limitation on the particle geometry or its constituent material that is being levitated, transported or rotated. For example, heavy tungsten spheres have been successfully levitated in air [11,12] and lightweight polystyrene particles were manipulated in 2D [13]. Another example includes the acoustic manipulation of liquid cylinders in fluid dynamics applications. Due to the squeezing effect of the radiation force when the sound field is activated, the liquid cylinder cross-section becomes elliptical[14] (Fig. 1). This effect is encountered in various applications involving the interaction of acoustical waves with highly compressible fluids and gas bubbles[9,15,16].

Numerical predictions and simulations for the acoustic radiation forces and torque on such particles, as well as the induced deformation are of particular importance in experimental design and optimization purposes. As reported in the scientific literature, significant works have been devoted to study some of the effects of ultrasonic progressive and standing waves [11,13,17-48] on particles. The shape of the particle being manipulated can range from an elongated cylindrical-like structure, such as carbon nano-tubes [49], elongated fibers [50], or a liquid bridge in air [14], to other oblate or prolate particles [51]. The modeling of the effect of the acoustic radiation force considered the infinitely-long *circular* geometry using standard analytical models[24,26-28,30,52-56] and numerical investigations [57] are clearly not suitable for non-circular geometries, especially when the particle's shape deviates strongly from its circular (unperturbed) one. Consequently, an improved theoretical formalism is needed that forms the basis of a reliable computational tool for the prediction of radiation forces and torque in the design of experimental applications involving non-circular particles.

The aim of this study is to develop a formal solution for the axial and transverse acoustic radiation force components as well as the radiation torque experienced by a sound-impenetrable elliptical cylinder placed arbitrarily in plane waves. The incident acoustic field is not restricted to progressive waves, so that the effects of (quasi)standing waves are also investigated. Furthermore, the analytical development is applicable to any frequency range. The procedure for evaluating the acoustic radiation force components and torque stems from an analysis of the acoustical scattering off the elliptical cylinder in a non-viscous fluid. Numerical simulations for the acoustic radiation force components and torque allow the predictions of potential scenarios of interest for various applications in fluid dynamics and particle manipulation.

Here, it is noted that an analytical model involving elliptical (modified) Mathieu functions has been previously developed [58] to evaluate the acoustic radiation force components and torque of plane progressive and standing waves on an elastic elliptical cylinder. However, the numerical implementation of the elliptical functions is not a straightforward task, since the angular wave functions are generally non-orthogonal. This difficulty may be resolved by adequate expansion of the angular Mathieu functions in terms of transcendental functions; nonetheless, this process is prone to possible errors (see for example the second column of Fig. 2 in Ref.[59], where the large resonances appear to be unphysical) and particular care should be given to check the adequate convergence of the series in the least-square sense and ensure that the wave-functions form a complete set on the surface of the elliptical cylinder. Furthermore, it has been mentioned in [58] "*that the acoustic radiation torque acting on the elliptic cylinder in a standing wave field vanishes due to the special symmetry of the wave fields.*" This statement as well as the lack of computational results for the radiation torque of standing waves in [58], seem to imply the vanishing of the torque for standing waves; this is actually the consequence of how the normal $\hat{\boldsymbol{n}}$ and transverse $\hat{\boldsymbol{t}}$ unit vectors have been inadequately defined in [58] with respect to the incident wavefront and not the surface of the object. It is important to emphasize that the mentioned statement (in Ref.[58]) is *only* valid for configurations dealing with end-on or broadside incidences of the incident field. Moreover, it should not be limited to standing waves. Note that the terminologies *end-on* or *broadside* incidences are defined when the direction of wave propagation of the incident beam coincides with either the larger or the smaller semi-axis of the ellipse, respectively. For example, assuming the semi-axis of the ellipse $a < b$, then the end-on incidence would correspond to $\alpha = \pi/2$, and the broadside incidence to $\alpha = 0$, as shown in Fig. 1 (where $\alpha$ is the angle of incidence). Contrary to the statement (in Ref.[58]) mentioned previously, it is obvious that the acoustic radiation torque acting on the elliptic cylinder in the field of progressive or (quasi)standing waves should *not* vanish when $0 < \alpha < \pi/2$. This has been also confirmed independently by experimental data [60,61].

In view of these developments, it is of some importance to develop an improved generalized formalism applicable to 2D objects with arbitrary cross-section, using standard Bessel and Hankel cylindrical wave functions widely available in standard computational packages. In the following, the partial-wave series expansion (PWSE) method is used, and the total field expression is forced to satisfy the Neumann boundary condition for a sound-impenetrable elliptical surface. The acoustic scattering problem of plane progressive, quasi-standing and standing waves by the elliptic cylinder is developed first, and then the derivations for the acoustic radiation force components and torque expressions would follow, based on an analysis of the far-field scattering. Tests for convergence and numerical results are illustrated and discussed. Additional perspectives and concluding remarks are presented as well.

## II. METHOD

Consider the case of plane quasi-standing (denoted in the following by the subscript *qst*) waves propagating in a nonviscous fluid, and incident upon an elliptical cylinder, forming an angle $\alpha$ with respect to the semi-axis $a$ that lies on the *x*-axis (see Fig. 1).

The incident field can be expanded in a partial-wave series expansion in a system of cylindrical coordinates $(r, \theta, z)$ with its origin chosen at the center of the ellipse as,

$$\Phi_{inc}^{qst} = \phi_0 e^{-i\omega t} \sum_n e^{in(\pi/2-\alpha)} \left[ e^{ikh} + R(-1)^n e^{-ikh} \right] J_n(kr) e^{in\theta}, \qquad (1)$$

where the summation $\sum_n = \sum_{n=-\infty}^{+\infty}$, $h$ is the distance in the axial direction from the center of the elliptical cylinder to the nearest velocity potential anti-node, $\alpha$ is the angle of incidence in the counter-clockwise direction as shown in Fig. 1, $R$ is the quasi-standing wave coefficient $0 \leq R \leq 1$ defined as the ratio $|\phi_1|/|\phi_0|$, where $\phi_0$ and $\phi_1$ are the amplitudes of the quasi-standing waves. $R = 1$ corresponds to the case of equi-amplitude "pure" standing waves, and $R = 0$ corresponds to the case of plane progressive waves. $J_n(\cdot)$ is the cylindrical Bessel function of first kind of order $n$, $\theta$ is the polar angle in the $(x,y)$ plane, and $k$ is the wave number of the incident radiation.

The scattered velocity potential field off the ellipse can be also represented using a PWSE as,

$$\Phi_{sca}^{qst} = \phi_0 e^{-i\omega t} \sum_n e^{in(\pi/2-\alpha)} \left[ e^{ikh} + R(-1)^n e^{-ikh} \right] C_n H_n^{(1)}(kr) e^{in\theta}, \qquad (2)$$

where $C_n$ are the scattering coefficient to be determined by applying the Neumann boundary condition for the sound-impenetrable rigid immovable surface with the assumption that the fluid surrounding the target is nonviscous, and $H_n^{(1)}(.)$ is the cylindrical Hankel function of the first kind, describing the propagation of outgoing waves in the surrounding fluid.

It is important to emphasize here that the cylindrical Hankel functions form a complete set suitable to represent the scattered waves on the surface of the ellipse as a result of analytic continuation [62]. The radius of convergence of the wave functions can be extended up to the object's boundary owing to Huygens' principle[63], which ensures adequate convergence of the partial-wave series as shown in the following (i.e. Figs. 2 and 3). Note, however, that this process may not be applicable to highly elongated ellipses or at high frequencies.

In 2D, the surface shape function $A_\theta$ of the ellipse depends on the polar angle $\theta$. The equation describing the elliptical surface shape function is given by,

$$A_\theta = \left[ (\cos\theta/a)^2 + (\sin\theta/b)^2 \right]^{-1/2}, \qquad (3)$$

where $a$ and $b$ are the semi-axes, respectively.

The core of the method is to apply the Neumann boundary condition [64] for the total (incident + scattered) steady-state velocity potential field at the surface of the elliptical cylinder, namely at $r = A_\theta$, such that,

$$\nabla\left(\Phi_{inc}^{qst}+\Phi_{sca}^{qst}\right)\cdot\mathbf{n}\Big|_{r=A_\theta}=0, \qquad (4)$$

where the normal vector **n** is expressed as,

$$\mathbf{n}=\mathbf{e}_r-\left(\frac{1}{A_\theta}\right)\frac{dA_\theta}{d\theta}\mathbf{e}_\theta, \qquad (5)$$

with $\mathbf{e}_r$ and $\mathbf{e}_\theta$ denoting the outward unit vectors along the radial and polar directions, respectively.

Substituting Eqs.(1) and (2) into Eqs.(4) using (5) leads to a system of linear equations,

$$\sum_n e^{in(\pi/2-\alpha)}\left[e^{ikh}+R(-1)^n e^{-ikh}\right]\left[\Gamma_n(\theta)+C_n\Upsilon_n(\theta)\right]=0, \qquad (6)$$

where the structural functions $\Gamma_n(\theta)$ and $\Upsilon_n(\theta)$ are expressed, respectively, as

$$\begin{Bmatrix}\Gamma_n(\theta)\\ \Upsilon_n(\theta)\end{Bmatrix}=e^{in\theta}\left[k\begin{Bmatrix}J_n'(kA_\theta)\\ H_n^{(1)'}(kA_\theta)\end{Bmatrix}-i\left(\frac{n}{A_\theta^2}\right)\frac{dA_\theta}{d\theta}\begin{Bmatrix}J_n(kA_\theta)\\ H_n^{(1)}(kA_\theta)\end{Bmatrix}\right], \qquad (7)$$

and the primes denote a derivative with respect to the argument.

Note that for a circular cylinder (i.e., $A_\theta=$ constant), the structural functions s $\Gamma_n(\theta)$ and $\Upsilon_n(\theta)$ are independent of the polar angle $\theta$, and thus, the scattering coefficients for the rigid circular cylinder can be recovered [64] as $C_n=-J_n'(ka)/H_n^{(1)'}(ka)$. In the case of an elliptical cylinder, however, the angular dependency should be eliminated in order to solve the system of linear equations for each partial-wave $n$ mode.

Subsequently, Eq.(6) is equated to a partial-wave Fourier series as,

$$\sum_n e^{in(\pi/2-\alpha)}\left[e^{ikh}+R(-1)^n e^{-ikh}\right]\left[\Gamma_n(\theta)+C_n\Upsilon_n(\theta)\right]=\sum_n[\psi_n+C_n\Omega_n]e^{in\theta} \qquad (8)$$
$$=0.$$

The coefficients $\psi_n$ and $\Omega_n$ are independent of the polar angle $\theta$, and they are determined after applying to Eq.(8) the following orthogonality condition,

$$\int_0^{2\pi}e^{i(n-\ell)\theta}d\theta=2\pi\delta_{n,\ell}, \qquad (9)$$

where $\delta_{i,j}$ is the Kronecker delta function.

This procedure leads to a new system of linear equations expressed as,

$$\sum_\ell[\psi_\ell+C_n\Omega_\ell]=0, \qquad (10)$$

where $\sum_\ell=\sum_{\ell=-\infty}^{+\infty}$, and

$$\left\{\begin{matrix}\psi_\ell \\ \Omega_\ell\end{matrix}\right\} = \frac{1}{2\pi}\sum_n e^{in(\pi/2-\alpha)}\left[e^{ikh}+R(-1)^n e^{-ikh}\right]\int_0^{2\pi}\left\{\begin{matrix}\Gamma_n(\theta) \\ \Upsilon_n(\theta)\end{matrix}\right\}e^{-i\ell\theta}d\theta. \qquad (11)$$

To obtain the coefficients $C_n$, Eq.(10) representing a coupled system, must be solved for each partial-wave mode number $n$ and summation index $\ell$. The integrals in Eq.(11) must be evaluated numerically before the summation can be carried out. Any suitable numerical integration method may be used, and the choice here is based on the Simpson's rule since it involves parabolic arcs (i.e., quadratic polynomials) instead of the straight lines (i.e., linear polynomials) used typically in the trapezoidal integration method. This method generally leads to insignificant numerical errors for smooth surfaces such as the elliptical cross-sectional area considered here. Once the coefficients $C_n$ are determined, evaluation of the scattering, radiation force and torque would follow based on the integration of second-order (quadratic) observables, as shown in the following.

Using the formulation of the scattering in the far-field [65], the radiation force expression can be derived based on the integration of the (Brillouin) radiation stress tensor over a surface at a large radius [66-68] enclosing the elliptical cylinder. Alternatively, one may choose to perform the integration on the surface of the elliptical cylinder using the near-field scattering. In a non-viscous fluid, these two approaches are commensurate with the same result [53,65] since the divergence of the radiation stress tensor is zero. Nevertheless, the use of the far-field scattering in the derivation of the radiation force and torque expressions presents an advantage because it requires fewer algebraic manipulations.

As such, the expression for the acoustic radiation force is given by [65]

$$\langle \mathbf{F}\rangle_{kr\to\infty} = \frac{1}{2}\rho k^2 \int_0^{2\pi}\text{Re}\{\Phi_{is}\}d\mathbf{S}, \qquad (12)$$

where $\text{Re}\{\cdot\}$ denotes the real part of a complex number, $\Phi_{is}\underset{kr\to\infty}{=}\Phi_{sca}^{qst*}\left[(i/k)\partial_r\Phi_{inc}^{qst}-\Phi_{inc}^{qst}-\Phi_{sca}^{qst}\right]$, the differential operator is $\partial_r = \partial/\partial r$, $d\mathbf{S}=dS\,\mathbf{e}_r$ where $dS = Lr\,d\theta$ is the scalar differential element of a cylindrical surface $S$ of length $L$ taken at a large radius $r$ enclosing the elliptical cylinder. The outward normal unit vector is $\mathbf{e}_r = \cos\theta\,\mathbf{e}_x + \sin\theta\,\mathbf{e}_y$, where $\mathbf{e}_x$ and $\mathbf{e}_y$ are the unit vectors in the Cartesian coordinates system, the symbol $\langle\cdot\rangle$ denotes time-averaging, and the superscript * denotes a complex conjugate.

For plane waves with arbitrary incidence, the elliptic cylinder would experience a transverse force in addition to the axial one. The axial (i.e. along the x-direction) and transverse (i.e. along the y-direction) radiation force components are defined as

$$\begin{aligned}\left\{\begin{matrix}F_x \\ F_y\end{matrix}\right\} &= \langle \mathbf{F}\rangle \cdot \left\{\begin{matrix}\mathbf{e}_x \\ \mathbf{e}_y\end{matrix}\right\} \\ &= \left\{\begin{matrix}Y_{qst,x} \\ Y_{qst,y}\end{matrix}\right\}S_c E_0,\end{aligned} \qquad (13)$$

where $S_c = 2bL$ is defined as the surface cross-section, $E_0 = \tfrac{1}{2}\rho k^2 |\phi_0|^2$ is a characteristic energy density, and $Y_{qst,\{x,y\}}$ are the non-dimensional radiation force function components.

Using the property of the angular integral,

$$\int_0^{2\pi} e^{i(n'-n)\theta} \begin{Bmatrix} \cos\theta \\ \sin\theta \end{Bmatrix} d\theta = \pi \begin{Bmatrix} (\delta_{n,n+1}+\delta_{n,n-1}) \\ i(\delta_{n,n+1}-\delta_{n,n-1}) \end{Bmatrix}, \qquad (14)$$

the substitution of Eqs.(1) and (2) [using the far-field limits of the Bessel and Hankel functions] into Eq.(12) using Eq.(14), leads to simplified forms for the axial and transverse radiation force function components written, respectively, as

$$Y_{qst,x} = \frac{1}{kb} \operatorname{Im} \sum_n \left\{ \begin{aligned} &(1+\alpha_n+i\beta_n)\left[(i\alpha_{n-1}+\beta_{n-1})e^{-i\alpha}+(i\alpha_{n+1}+\beta_{n+1})e^{i\alpha}\right] \\ &\times\left[(R^2-1)+2i(-1)^n R\sin(2kh)\right] \end{aligned} \right\}, \qquad (15)$$

$$Y_{qst,y} = -\frac{1}{kb} \operatorname{Re} \sum_n \left\{ \begin{aligned} &(1+\alpha_n+i\beta_n)\left[(i\alpha_{n+1}+\beta_{n+1})e^{i\alpha}-(i\alpha_{n-1}+\beta_{n-1})e^{-i\alpha}\right] \\ &\times\left[(R^2-1)+2i(-1)^n R\sin(2kh)\right] \end{aligned} \right\}, \qquad (16)$$

where $\operatorname{Im}\{\cdot\}$ denotes the imaginary part of a complex number, and the coefficients $\alpha_n$ and $\beta_n$ are the real and imaginary parts of the coefficients $C_n$ ($= \alpha_n + i\beta_n$).

When $\alpha = 0$, the transverse component $Y_{qst,y}$ vanishes, which corresponds to the axial (or on-axis) case. This may be anticipated from Eq.(16) by noticing that $\sum_n C^*_{n+1} = \sum_n C^*_{n-1}$ (or alternatively $\sum_n C_{n+1} = \sum_n C_{n-1}$).

Similarly, when $\alpha = \pi/2$, the axial component $Y_{qst,x}$ [given in Eq.(15)] vanishes.

From Eqs. (15) and (16), the radiation force components for the progressive $Y_{p,\{x,y\}}$ and standing wave $Y_{st,\{x,y\}}$ cases can be deduced by taking $R = 0$ or 1, respectively.

The analysis of the far-field scattering is extended to derive the acoustic radiation torque expression. Based on the conservation law of angular momentum, the time-averaged acoustic radiation torque is evaluated by integrating the moment of the time-averaged (Brillouin) radiation stress tensor over a cylindrical surface in the far-field enclosing the elliptical cylinder. Taking the expression given by Eq.(10) in Ref.[69] in the far-field, the acoustic radiation torque can be expressed as,

$$\langle \mathbf{N} \rangle \underset{kr\to\infty}{=} -\rho \iint_S \langle \mathbf{v}(\mathbf{r}\times\mathbf{v}) \rangle \cdot d\mathbf{S}, \qquad (17)$$

where vector velocity is $\mathbf{v} \underset{kr\to\infty}{=} \nabla\Phi \underset{kr\to\infty}{=} \nabla(\Phi_{inc} + \Phi_{sca})$, and $\mathbf{r}$ is the radius vector.

Using the properties for the time-average of the product of two complex numbers, the only non-vanishing component (i.e. in the direction $z$ perpendicular to the polar plane) of the acoustic radiation torque can be rewritten in terms of the total (incident + scattered) velocity potential field as,

$$N_{qst,z}^{rad} = \langle \mathbf{N} \rangle \cdot \mathbf{e}_z,$$
$$= \frac{\rho}{2} \text{Im} \left\{ \iint_S (\partial_r \Phi^*) \hat{L}_z \Phi \, dS \right\}, \quad (18)$$

where $\mathbf{e}_z$ is the unitary vector along the z-direction, and $\hat{L}_z$ is the z-component of the angular momentum operator in polar coordinates given by

$$\hat{L}_z = -i \frac{\partial}{\partial \theta}. \quad (19)$$

After taking the far-field limits of Eqs.(1) and (2), and substituting them into Eq.(18) using Eq.(19), the expression for the acoustic radiation torque component can be obtained after some algebraic manipulation using the property of the following angular integral,

$$\int_0^{2\pi} e^{i(n'-n)\theta} d\theta = 2\pi \delta_{n,n'}. \quad (20)$$

Subsequently, the expression for the acoustic radiation torque (per-length) is obtained as,

$$N_{qst,z}^{rad}/L = -2\rho |\phi_0|^2 \sum_n n \left[ (R^2+1) + 2(-1)^n R\cos(2kh) \right] \left[ \alpha_n + \alpha_n^2 + \beta_n^2 \right]. \quad (21)$$

For computational purposes, it is convenient to define a non-dimensional radiation torque function such that, $\tau_{qst,z} = N_{qst,z}^{rad}/(E_0 \pi b^2 L)$. Its expression is given as,

$$\tau_{qst,z} = -\frac{4}{\pi(kb)^2} \sum_n n \left[ (R^2+1) + 2(-1)^n R\cos(2kh) \right] \left[ \alpha_n + \alpha_n^2 + \beta_n^2 \right]. \quad (22)$$

It is particularly important to note from Eq.(22) that the monopole partial-wave term ($n = 0$) does not contribute to the torque generation. Moreover, the term $\left[ \alpha_n + \alpha_n^2 + \beta_n^2 \right]$ in Eq.(22) is equal to $\text{Re}\{C_n\} + |C_n|^2$. This factor appears explicitly in the expression of the absorption cross-section $\sigma_{abs}$ (or the absorption efficiency $Q_{abs}$) [70,71], which vanishes for the case of a non-absorptive *circular cylinder*. That is, a perfectly rigid, non-viscous fluid-filled or elastic cylinder of *circular* cross-section (in arbitrary wavefronts) does not experience any radiation torque since $\text{Re}\{C_n\} = -|C_n|^2$. This property, however, does *not* hold for the rigid *elliptical cylinder* because $\text{Re}\{C_n\}|_{\text{elliptical cylinder}} \neq -|C_n|^2|_{\text{elliptical cylinder}}$.

When the angle of incidence $\alpha = 0$ or $\pi/2$, the radiation torque on the rigid elliptical cylinder vanishes as required by symmetry. Though the expressions given by Eqs. (21) or (22) do not display an explicit dependence on $\alpha$, it should be emphasized that this parameter is intrinsically coupled to the functions $\psi_\ell$ and $\Omega_\ell$ [as shown in Eq.(11)] used to calculate the scattering coefficients $C_n$. In other words, the coefficients $C_n$ (or alternatively $\alpha_n$ and $\beta_n$) incorporate the angle dependency effects in them.

### III. NUMERICAL RESULTS AND DISCUSSION

The analysis is started by performing adequate checks for convergence of the radiation force and torque expressions at various incidence angles $\alpha$. These tests are crucial in order to verify the accuracy of the results and validate the PWSE method.

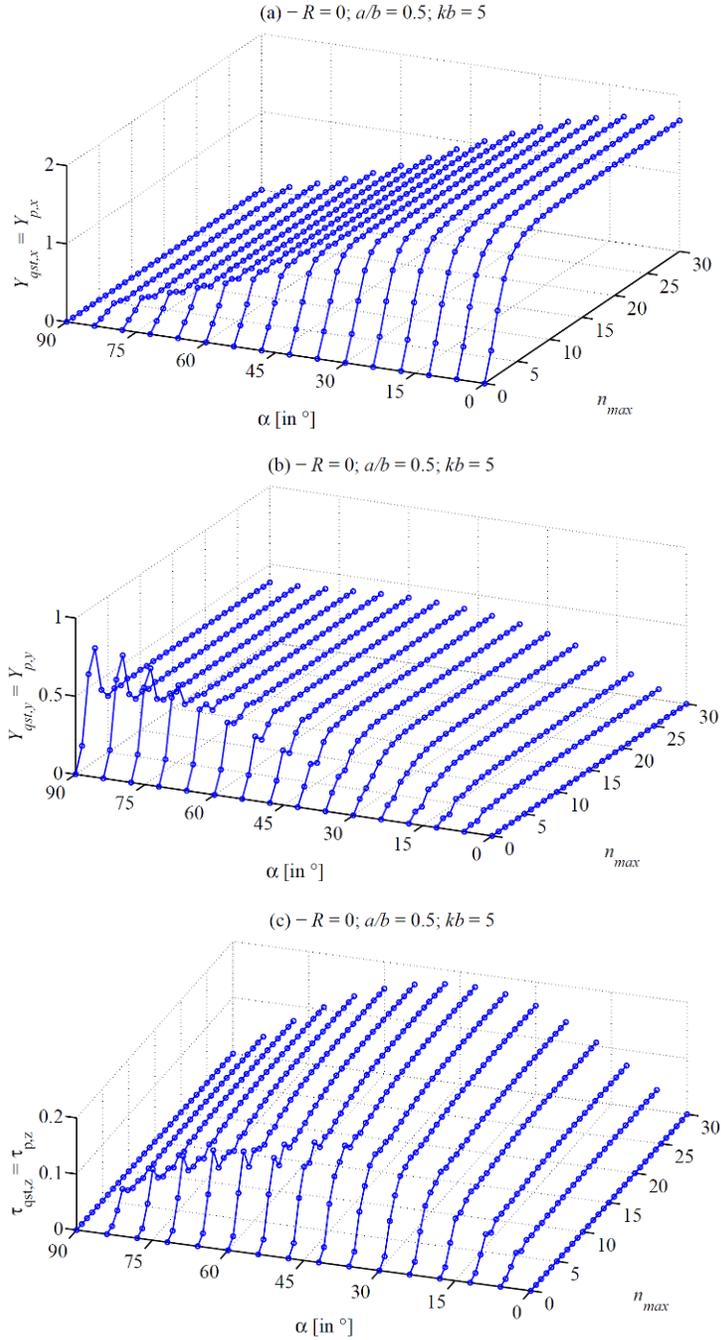

FIG. 2. Panels (a) and (b) show the convergence plots for the axial and transverse radiation force functions, respectively, for a rigid elliptical cylinder with an aspect ratio $a/b = 0.5$ at $kb = 5$ in plane progressive waves ($R = 0$). The angle $\alpha$ ranges from broadside ($\alpha = 0°$) to end-on ($\alpha = 90°$) incidence in an incremental step of $\delta\alpha = 6°$, and $0 \leq n_{max} \leq 30$. Panel (c) displays the convergence plot for the radiation torque function. The convergence to the stable solution for the axial and transverse radiation force functions requires about 15 terms in the series to ensure adequate convergence, while for the radiation torque function, about 25 terms need be included so that the radiation torque expression converges to the steady limit with minimal truncation error.

The coupled system given by Eq.(10) must be solved by matrix inversion procedures after a specific truncation limit is adequately defined and applied to the series. Written in matrix notation and performing appropriate inversion, the solution of Eq.(10) for the scattering coefficients is obtained as[72], $\mathbf{C} = -\mathbf{\Omega}^{-1}\mathbf{\psi}$.

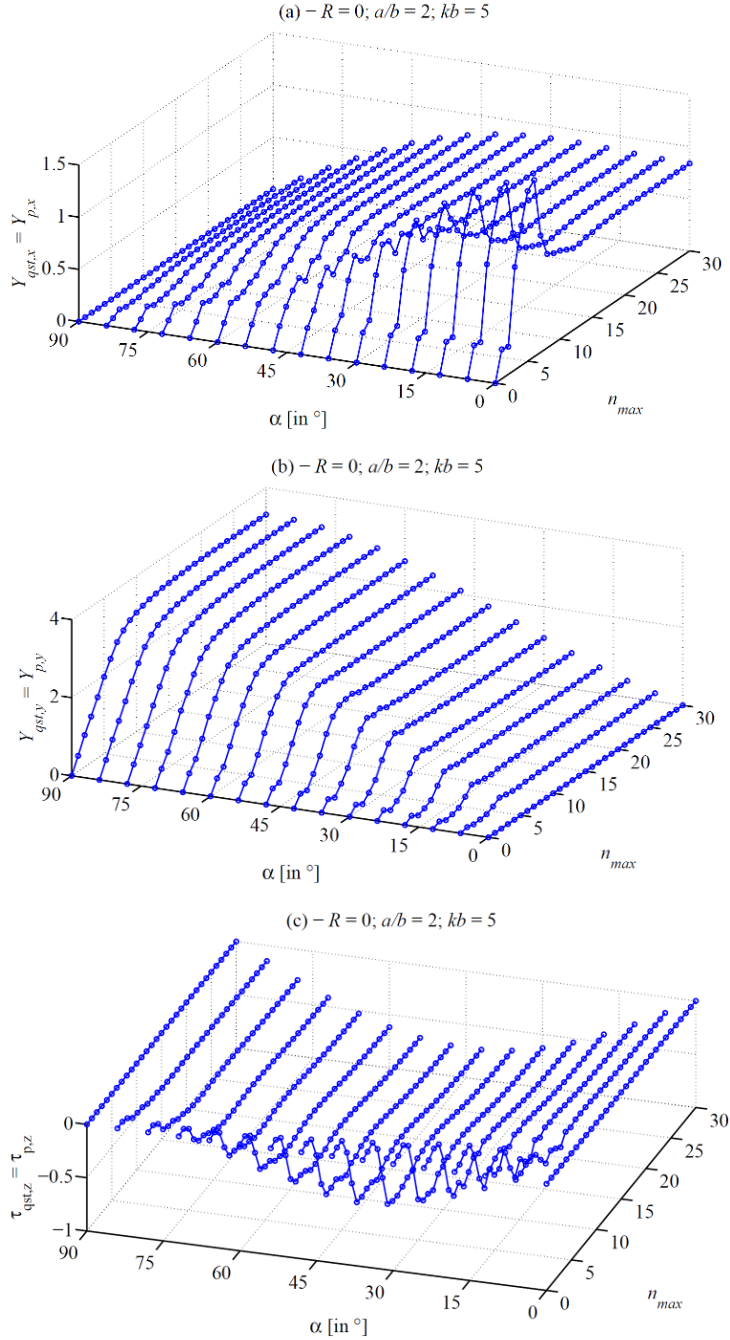

FIG. 3. The same as in Fig. 2, but for a rigid elliptical cylinder with an aspect ratio $a/b = 2$ at $kb = 5$ in plane progressive waves ($R = 0$). The angle $\alpha$ ranges from end-on ($\alpha = 0°$) to broadside ($\alpha = 90°$) incidence, and $0 \leq n_{max} \leq 30$. For this aspect ratio, the convergence to the stable solution for the axial and transverse radiation force functions requires about 25 terms in the series to ensure adequate convergence, while for the radiation torque function, about 28 terms need be included so that the radiation torque expression converges to the steady limit with minimal truncation error.

Practically, Eq. (10) is solvable satisfactorily, provided that the convergence is satisfied by continually increasing the maximum truncation limits $\ell_{max}$ and $n_{max}$. If $(\ell_{max}, n_{max})$ are generally smaller than the non-dimensional size parameter $ka$ (or $kb$), the results obtained from $C_n$ will be inaccurate. On the other hand, if $(\ell_{max}, n_{max})$ are too large so that many terms beyond what is required for convergence are included, at best the calculation will be needlessly time-consuming. At worst, numerical instabilities can arise due to numerical round-off errors and loss of accuracy in computing the cylindrical wave functions with large order. Therefore an optimal truncation scheme is required so that suitable values for $(\ell_{max}, n_{max})$ are selected so as to achieve adequate convergence. An adequate truncation limit is defined based on a convergence criterion such that $|C_{n+n_{max}}/C_0| \sim 10^{-16}$. This criterion warrants a negligible truncation error, as shown in Figs. 2 and 3 for an elliptic cylinder having an aspect ratio $a/b$ = 0.5 and 2, respectively. It is interesting to note from these figures the vanishing of the axial and transverse radiation force functions as well as the radiation torque function for the monopole ($n$ = 0) term of the elliptical cylinder. These convergence plots further demonstrate the adequate validation of the analytical formalism presented here throughout the PWSE method.

Now, the behavior of the solutions is illustrated for the axial and transverse radiation force functions given by Eqs.(15) and (16), respectively, as well as the radiation torque function given by Eq.(22). Numerical examples are computed by developing a MATLAB code that involves solving the coupled system of linear equations given by Eq.(10) for each partial-wave index number $n$ with adequate numerical integration of Eq.(11) using Simpson's rule, then perform the matrix inversion to determine the coefficients $C_n$. in the simulations, the parameter $h$ is chosen to be dependent on the wavenumber $k$ such that $h = \pi/(4k)$. Moreover, the non-dimensional radiation force and torque functions are evaluated in the bandwidths $0 < kb \leq 5$, $0° \leq \alpha \leq 90°$ with particular emphasis on varying the reflection coefficient $R$ showing the transition from the progressive ($R$ = 0) to the standing ($R$ = 1) wave case, and the aspect ratio $a/b$ of the elliptical cylinder. The surrounding fluid is chosen to be (non-viscous) water with its assumed physical properties, $\rho$ (mass density) = 1000 kg/m$^3$, $c$ (speed of sound) = 1500 m/s.

The first set of computational plots considers the numerical evaluation of the radiation force functions given by Eqs.(15) and (16), respectively, as well as the radiation torque function given by Eq.(22) for progressive waves with $R$ = 0. For a specific value of the aspect ratio $a/b$, the results are displayed in each row of Fig. 4 for $Y_{p,x}$, $Y_{p,y}$ and $\tau_{p,z}$ (where the subscript $p$ represents progressive waves). The left column of Fig. 4 corresponding to panels (a),(d),(g),(j) and (m) displays the $Y_{p,x}$ plots for $a/b$ = 0.5, 0.75, 1, 1.5 and 2, respectively. As $kb$ increases and for $\alpha$ = 0°, $Y_{p,x}$ is maximal and grows monotonically before reaching a maximum peak that decreases as the aspect ratio $a/b$ increases. As $\alpha$ increases, $Y_{p,x}$ reduces in amplitude for $a/b \leq 1$, while it increases for $a/b > 1$ before vanishing for $\alpha$ = 90°. The exact opposite situation is encountered for the transverse component; for $\alpha$ = 0°, $Y_{p,y}$ vanishes as required by symmetry. As $kb$ increases and for $\alpha > 0°$, $Y_{p,y}$ grows monotonically before reaching a maximum peak then a plateau. Its amplitude increases as the aspect ratio $a/b$ increases, and reaches a maximum for $\alpha$ = 90° as shown in panels (b),(e),(h),(k) and (n) of the central column of Fig. 4. Note also that for $a/b$ = 1, panels (g) and (h) are exactly anti-symmetric. As for the radiation torque function, the plots in panels (c),(f),(i),(l) and (o) display the variations of $\tau_{p,z}$ versus $kb$ and $\alpha$. For $a/b < 1$, the torque function is positive, meaning that the sense of rotation is in the counter-clockwise direction, whereas for $a/b > 1$, the torque function reverses sign, meaning that the sense of rotation is in the clockwise direction. The torque function vanishes for the circular cylinder (i.e., $a/b$ = 1) $\forall$ $\alpha$. Furthermore, $\tau_{p,z}$ vanishes for $\alpha$ = 0° or 90° for all the plots. Only the position at $\alpha$ = 90° will be a stable location for the elliptical cylinder because $d\tau_{p,z}/d\alpha < 0$ close to that angle, and the torque is always directed towards that position. Note also that the torque function becomes larger (in the absolute sense) as $a/b$ increases, and its amplitude is maximal at $\alpha \approx 45°$.

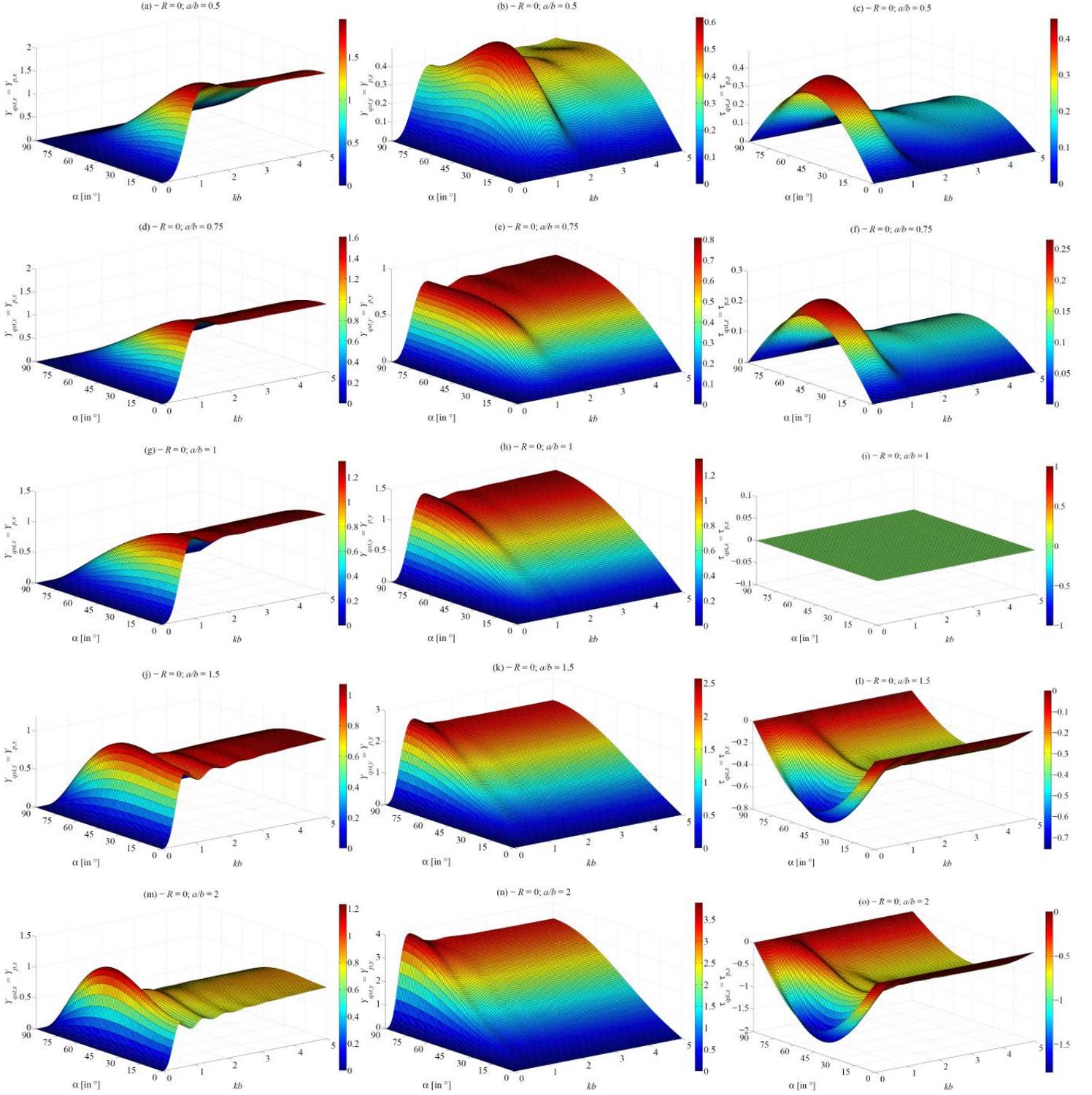

FIG. 4. The left and middle columns display the axial and transverse acoustic radiation force function plots for a rigid elliptical cylinder in plane progressive waves ($R = 0$), versus $kb$ and $\alpha$. The panels in the right column show the radiation torque function plots. The change in the aspect ratio $a/b$ of the rigid elliptical cylinder has a significant effect of the amplitude of the radiation force and torque, as well as the direction of rotation.

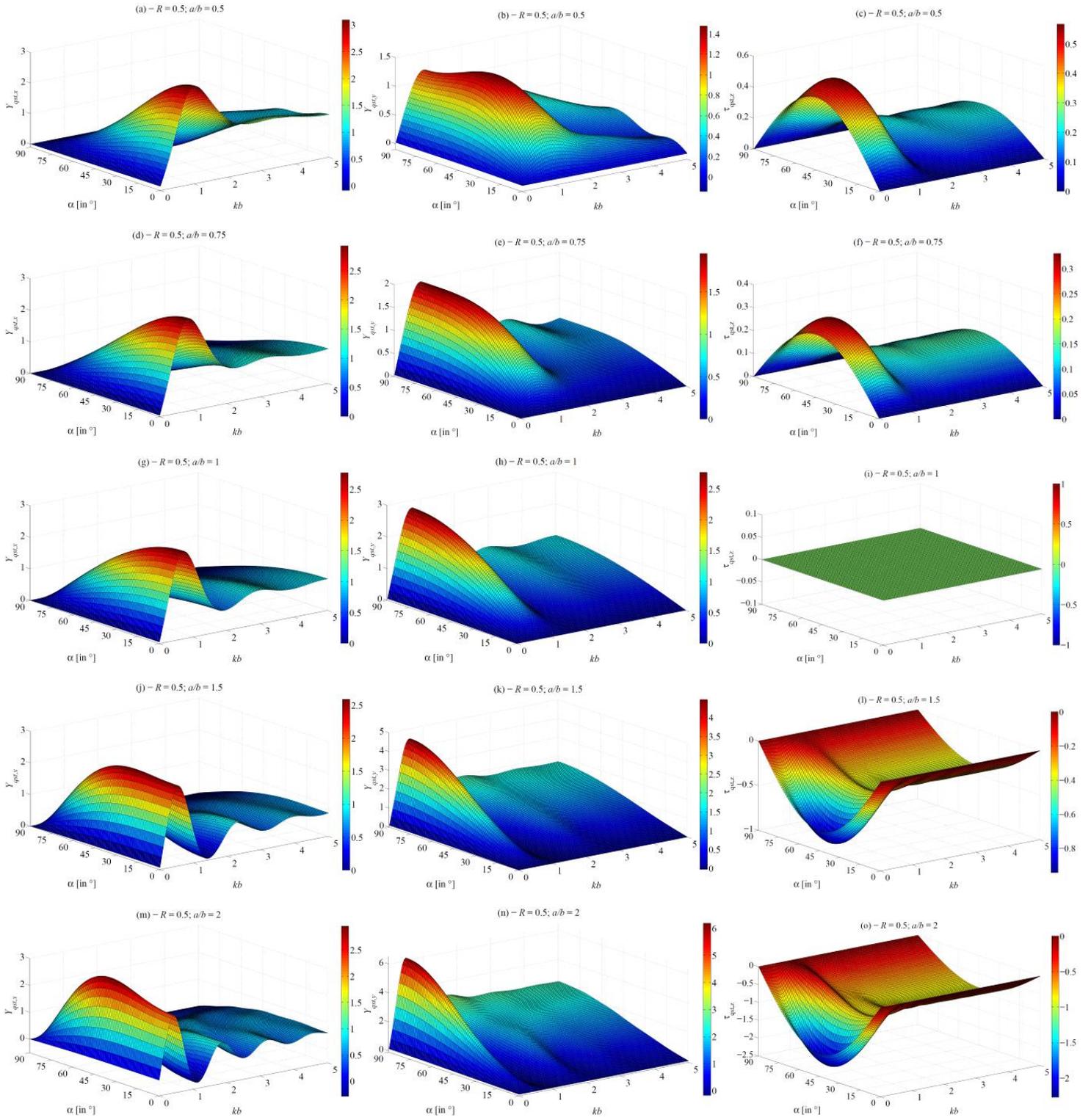

FIG. 5. The same as in Fig. 4 but the incident field corresponds to quasi-standing waves with $R = 0.5$.

The effect of increasing the reflection coefficient $R = 0.5$ is shown in the panels of reflected from the edge of the ellipse [73], which interfere with circumferential waves (known as Franz waves) propagating in the exterior fluid surrounding the elliptical cylinder. The modification in the path length for the Franz waves circumnavigating the ellipse occurs as the aspect ratio of the elliptical cylinder changes, as well as their phase velocity [73].

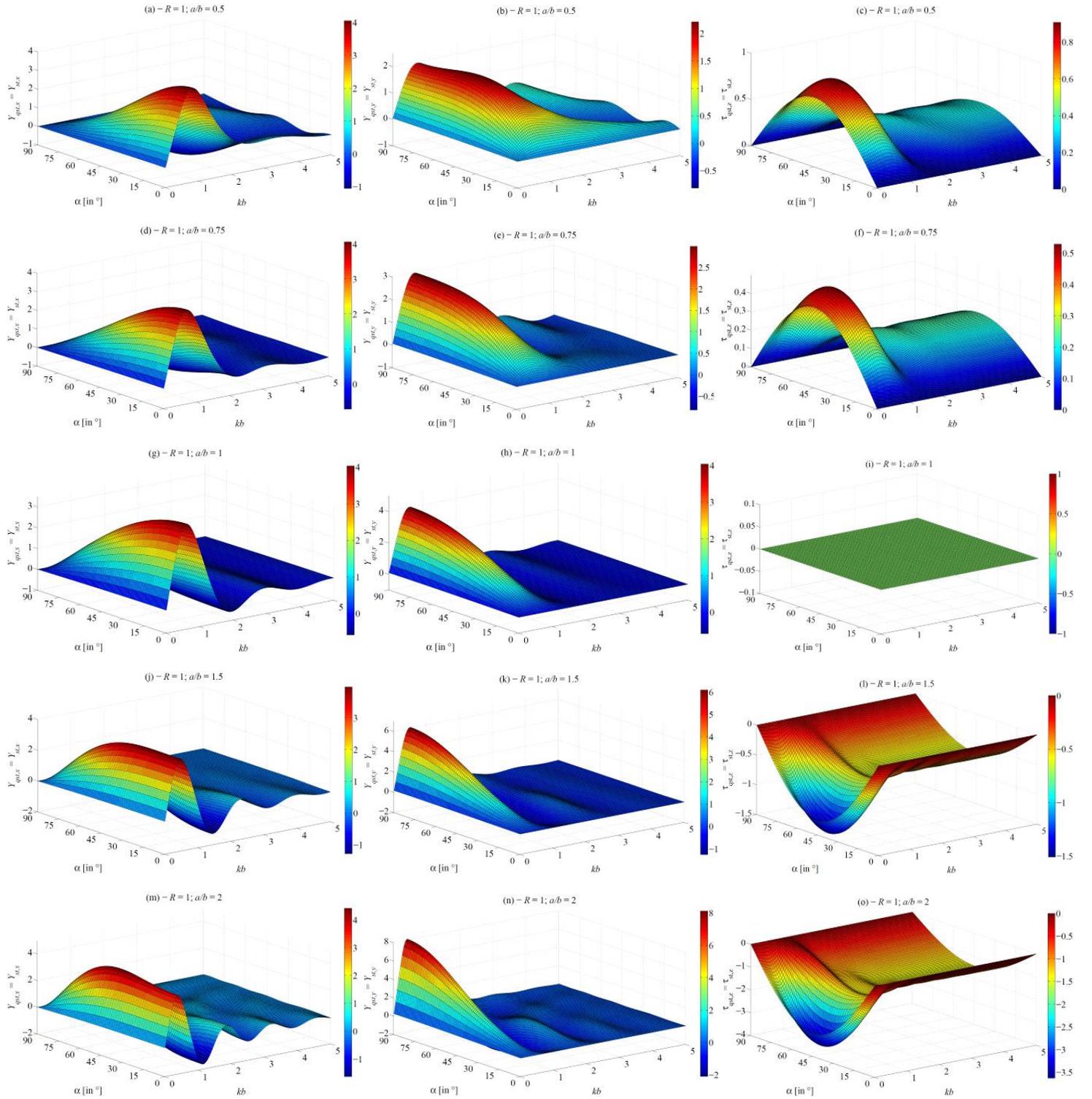

FIG. 6. The same as in Fig. 4 but the incident field corresponds to standing waves with $R = 1$.

Fig.5, which corresponds to the case of quasi-standing waves. The one-to-one comparison with each panel of Fig. 4, shows that the plots for $Y_{qst,\{x,y\}}$ and $\tau_{qst,z}$ are larger for the quasi-standing wave case as the acoustic field is stronger in amplitude (or intensity). One also notices that the frequency of the oscillations/undulations in the plots increases as the aspect ratio *a/b* increases. These oscillations are the result of the waves that are specularly

With the additional increase of the reflection coefficient to $R = 1$, which corresponds to the case of standing waves (denoted by the subscript *st*), larger amplitudes for the radiation force and torque functions are obtained, as shown in panels (a)-(o) of Fig. 6. As shown in the radiation force function plots (i.e., the panels of the left and central columns), there exist specific values determined by an appropriate choice of *kb* and $\alpha$ for which $Y_{st}$ is negative. In this case, the radiation force exerted on the elliptical cylinder is directed toward a pressure antinode of the standing wave field. Moreover, by comparing the plots with those of Fig. 5, it is clear that the oscillations/undulations are more manifested and further enhanced, especially when *a/b* > 1. As *kb* increases, $Y_{st}$ oscillates around zero. As also noted previously for the progressive wave case, the torque functions in (quasi)standing waves $\tau_{qst,z}$ and $\tau_{st,z}$ vanish for the circular cylinder (i.e., *a/b* = 1) $\forall\ \alpha$, and for $\alpha = 0°$ or 90° for all the plots. The plots also show that the radiation torque is less affected by the nature of the incident field (i.e., progressive versus standing waves), in comparison to the radiation force function plots that are very sensitive to the type of incoming waves.

## IV. CONCLUDING REMARKS

In this contribution, a formal theoretical analysis augmented by numerical computations for the axial and transverse acoustic radiation forces and torque of plane progressive, quasi-standing and standing waves with arbitrary incidence upon a 2D rigid elliptical (non-circular) cylinder is developed. The present benchmark analytical solution could be used to validate results obtained by strictly numerical methods, such that the standard finite element method (FEM) or other. In the simulations, particular emphasis is given on varying the reflection coefficient *R* in order to demonstrate the transition from the progressive to the standing wave case. Moreover, the choice for the angle of incidence and aspect ratio are important parameters that should be taken into account for optimal experimental design. The results reveal that the elliptical geometry has a direct influence on the radiation force and torque, so that the classical theory for circular cylinders (at normal incidence) leads to significant miscalculations when the cylinder cross-section becomes non-circular. The elliptical cylinder experiences, in addition to the acoustic radiation force, a radiation torque that sets the particle into rotation and vanishes for the circular cylinder. Depending on the aspect ratio, the torque reverses sign, meaning that the elliptical cylinder may rotate in the counter-clockwise or clockwise direction depending on the choice of the aspect ratio *a/b*.

At the same time, it is essential to mention that the present analysis has been limited to moderately flat and elongated cross-sections, so that an aspect ratio of 3:1 has not been exceeded. For highly flat or elongated elliptical particles (for example compressible cylindrical air bubbles) exceeding this aspect ratio of 3:1, the numerical implementation using the PWSE requires some improvements so as the convergence of the coupled system of linear equations [i.e., Eq.(10)] is warranted. For surfaces strongly deviating from the moderately elongated geometry, Eq.(10) may become unsolvable because of an ill-conditioning [74] during matrix inversion procedures. This instability arises as a consequence of taking a large number of cylindrical partial-waves of order *n* to fit a non-circular object. This difficulty may be also encountered for the case where the wave frequency increases, as more terms in the series are needed to ensure adequate convergence. Fortunately, there exist several methods and techniques [75,76] that could be used to advantage in order to improve the present formalism for highly elongated (or extremely flat) surfaces, such as the iterative Gram-Schmidt orthogonalization procedure [77], a boundary condition enforcement by point-matching [78], the use of extended-precision floating-point variables [79], or other methods. This ill-conditioning should nonetheless not be interpreted as a lack of rigor of the PWSE method, rather this technique must be further improved and conditioned for highly elongated or extremely flat objects. Applying such improvements to the method, other irregularly-shaped 2D geometries may be considered to investigate the radiation force and torque effects. Such geometries may include a Chebyshev surface [80-82], a square [83], an oval or a stadium [84], a super-ellipse, or other. Note that the 3D spheroid case has been recently investigated from the standpoint of acoustic scattering [85,86] and radiation force[87,88] theories.

In this work, the assumption of an ideal (non-viscous) fluid surrounding the elliptical cylinder is considered. Nonetheless, it must be emphasized that the derived equations for the axial and transverse radiation force and torque functions [given by Eqs.(15),(16) and Eqs.(21) or (22)] remain valid for a thermoviscous fluid, provided that the penetration depths of viscous and thermal waves are much smaller than the acoustic wavelength of the incident radiation, as well as the dimensions of the particle [89]. Yet, the scattering coefficients of the elliptical cylinder in a thermoviscous fluid must be determined using standard methods of continuum mechanics based on the continuity and Navier-Stokes equations [90,91], so that the corresponding coefficients (equivalent to $\alpha_n$ and $\beta_n$) would incorporate the thermoviscous effects in them prior to their use in Eqs.(15),(16) and Eqs.(21) or (22).